# Intrinsically shunted Josephson junctions for electronics applications

M. Belogolovskii[1,2], E. Zhitlukhina[2,3], V. Lacquaniti[4], N. De Leo[4], M. Fretto[4], A. Sosso[4]

[1]*Institute for Metal Physics, Kyiv, 03142, Ukraine*
E-mail: belogolovskii@ukr.net
[2]*Donetsk National University, Vinnytsia, 021021, Ukraine*
[3]*Donetsk Institute for Physics and Engineering, Kyiv, 03028, Ukraine*
[4]*National Institute for Metrological Research, Torino, 10135, Italy*

**Abstract**

Conventional Josephson metal-insulator-metal devices are inherently underdamped and exhibit hysteretic current-voltage response due to a very high subgap resistance compared to that in the normal state. At the same time, overdamped junctions with single-valued characteristics are needed for most superconducting digital applications. The usual way to overcome the hysteretic behavior is to place an external low-resistance normal-metal shunt in parallel with each junction. Unfortunately, such solution results in a considerable complication of the circuitry design and introduces parasitic inductance through the junction. This paper provides a concise overview of some generic approaches that have been proposed in order to realize internal shunting in Josephson heterostructures with a barrier that itself contains the desired resistive component. The main attention is paid to self-shunted devices with local weak-link transmission probabilities so strongly disordered in the interface plane that transmission probabilities are tiny for the main part of the transition region between two superconducting electrodes, while a small part of the interface is well transparent. We consider the possibility of realizing a universal bimodal distribution function and emphasize advantages of such junctions that can be considered as a new class of self-shunted Josephson devices promising for practical applications in superconducting electronics operating at 4.2 K.

1. INTRODUCTION

Modern integrated circuits are based on a silicon complementary metal-oxide-semiconductor (CMOS) technique that uses a combination of *p*- and *n*-type metal-oxide-semiconductor field-effect transistors to implement logic gates and other digital circuits. Shrinking the size of the field-effect transistors has improved the functionality, speed, and the cost of microprocessors over the last four decades. However, the advantages of scaling are quickly going down and, in particular, the operational frequency of central processing units has stopped its improving since 2003 due to power consumption of the circuitry reaching their cooling limit ($\approx$100 W/cm$^2$) [1-2]. With the end of the Moore's Law in sight, various technologies and computing models are actively proposed as potential alternative to CMOS circuits.

Superconducting digital electronics may be an attractive candidate for the replacement of the semiconductor technology with many potential benefits [3-5]. Its most popular version known as single-flux quantum (SFQ) logic [6] is using Josephson junctions as ultrafast switches and magnetic-flux encoding of information. It offers a combination of high speed and very low energy consumption and thus allows to develop small-scale computational circuits that dissipate more than one thousand times less power than state-of-the-art silicon SMOS circuits [5]. Certainly, from a future perspective, we should also include in the estimations the energy required for the refrigeration needs. But modern, closed-cycle cryocoolers can already provide cooling up to 4 K without need of liquid helium. Even taking into account their energy consumption the effective SFQ switching energy remains to be an order of magnitude lower than state-of-the-art CMOS [5]. That is why this technology which requires cryogenic refrigeration can be energy efficient compared to existing room-temperature semiconductor circuits.

In 2014, "The Next Wave", the USA National Security Agency's review on emerging technologies published a foresight report on the actual state and prospective development of superconductive electronics [5]. As was emphasized in Ref. 5, two promising research directions are urgent for a further progress of Josephson-junction circuitry. The first one relates self-shunted devices able to eliminate the need for external-shunt resistors and the second one concerns hybrid structures with ferromagnetic interlayers which could realize junctions with a built-in π phase shift of the superconductive wave function as well as create a SFQ-compatible memory element [7].

Our review is devoted just to the first problem, namely, possible ways towards creating internally shunted Josephson multilayered structures. In the second section of the paper, we discuss what characteristics of individual Josephson junctions are required by superconductive applications. The third section highlights some previously proposed solutions for internally shunted devices without need of an external resistor. Theoretical background and experimental verification of the self-shunting effect in Josephson devices with strongly disordered barriers is the subject of the fourth section where it is related to specific features of a multichannel inhomogeneous weak link between superconductive electrodes. We present a summary of our results for two types of intrinsically shunted Josephson junctions with a self-averaging distribution of weak-link transparencies. The Conclusions provide some comments relating applications of the intrinsically shunted junctions in superconductive circuitries and offers few concluding remarks.

## 2. REQUIRED CHARACTERISTICS OF JOSEPHSON JUNCTIONS FOR SUPERCONDUCTIVE CIRCUIT TECHNOLOGY

We start with the discussion how planar superconductor-weak link-superconductor junctions with lateral dimensions smaller than the Josephson penetration depth [8] can be characterized and what parameters are needed for superconducting circuits. The most stable technological process used now in different analog and digital low-temperature superconductor applications, including very-large-scale SFQ circuits, is based on the all-refractory Nb-AlO$_x$-Nb planar thin-film technology [9,10]. It allows the fabrication of circuits with up to tens of thousands of Josephson junctions and consists of the deposition of three films, superconducting (S) niobium base and counter electrodes and an ultra-thin aluminum interlayer, oxidized to create an aluminum-oxide insulating (I) interlayer. The I interlayer plays the role of a tunnel barrier whose thickness, about 1 nm, determines the critical current density $J_c$ of the device. In such SIS samples the current-vs-Josephson phase relation is known to be sinusoidal [8]. The main part of our discussion below remains within the Nb-based technology and will be focused mostly on the electronics operation at 4.2 K, the liquid-helium temperature that makes possible to achieve ultralow switching energy in the SFQ devices [5].

*Current-voltage curves.* SFQ logic requires Josephson junctions with non-hysteretic current-voltage (*I–V*) characteristics while conventional SIS devices are inherently underdamped and exhibit hysteretic *I-V* response, see Fig. 1a. The usual way to overcome the hysteretic behavior is to place an external low-resistance normal-metal shunt in parallel with each junction [5]. It permits to achieve the required damping and well-behaved SFQ pulse generation but, at once, results in a considerable complication of the circuitry design due to additional wiring and introduces significant parasitic inductance through the device which limits its high frequency operation. Moreover, the parasitic inductance in shunted Josephson SIS trilayers can lead to stable negative-resistance regions in the *I-V* curves and a strongly increased low-frequency voltage noise arising from chaotic transients between subharmonic modes [11]. Thus, the elimination of the majority of external shunt resistors by using self-shunted superconducting junctions can be regarded as an important task for further progress of the superconductive Josephson circuitry.

Properties of an individual Josephson junction can be simulated by a RCSJ (Resistively and Capacitively Shunted Junction) model [8] with the quasiparticle resistance *R* in the operating voltage range that can strongly depend on *V* in many cases, the capacitance C, and the critical supercurrent $I_c$ in parallel (Fig. 1b). In ideal SIS trilayers with infinitely small barrier transparencies $D \ll 1$ quasiparticle transport at voltages $V < 2\Delta/e$ (the subgap regime), where $\Delta$ is the superconducting energy gap, is possible only due to thermally activated processes and, hence, rather high subgap resistances $R_{sg}$ are realized. For voltages above the gap value, however, direct tunneling of quasiparticles from one superconducting electrode to unoccupied states on the other side of the barrier is allowed and with increasing *V*, the device resistance is quickly approaching the normal-state value $R_N$.

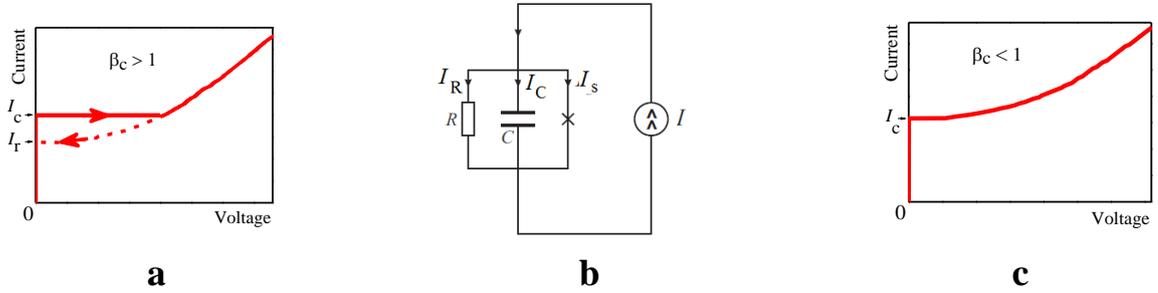

*Fig. 1*. Hysteretic, typical for an SIS trilayer (*a*) and single-valued, realized in an SNS sample, (*c*) current-voltage characteristics at *T* = 0 as well as the basic RCSJ equivalent circuit (*b*) for a current-biased Josephson junction.

Within the RCSJ model (Fig. 1b), the Josephson-junction operation regime (if it is hysteretic or not) is characterized by the McCumber-Stewart damping parameter $\beta_c$, a product of the characteristic Josephson frequency $\omega_c = 2eV_c/\hbar$, where $V_c = I_c R$, and the decay time $\tau_{RC} = RC$

$$\beta_c = \frac{2eV_c RC}{\hbar}. \tag{1}$$

Nonhysteretic *I-V* behavior is achieved when $\beta_c$ is less than unity. When the practically used voltage range in a non-shunted SIS junction is limited by the gap value, *R* being a very high subgap resistance $R_{sg}$ (its value can be found experimentally by applying magnetic fields and reducing the critical current to zero) and $\beta_c = 2eI_c R_{sg}^2 C/\hbar \gg 1$. In the next section, we discuss how the latter parameter can be suppressed below unity and, as a result, the current-voltage curve will be single-valued (Fig. 1c).

*Superconducting critical current.* In different Josephson-junction configurations, the critical current $I_c$ scales as the inverse of the normal-state resistance $R_N$ [12] and thus, the $I_c R_N$ product should not depend on the device dimensions and can be considered an important figure of merit for SFQ circuitry components that determines the strength of Cooper pair tunneling and therefore should be as large as possible. There is also another argument for increasing the $I_c R_N$ value. It is inversely proportional to the SFQ-pulse width [5] and, hence, junctions with high $I_c R_N$ products will potentially increase circuit speeds. From the other hand, the energy dissipated for each basic SFQ switching event is proportional to $I_c$. Thus, decreasing $I_c$ could result in exceptionally low switching energy. In order to avoid high error rates, it is important to operate with switching energies several orders of magnitude higher than the thermal energy at 4.2 K [5]. So, the optimal choice of the critical supercurrent value should be found as a compromise between different requirements. Note that the critical current density $j_c$ of the Josephson junction based on conventional superconductors with a comparatively low critical temperature $T_c$, a major issue for both fabrication and application, is set during the fabrication process by the oxygen exposure dose, the product of the oxygen partial pressure and the oxidation time [13, p. 285].

Zero-temperature value of the product $I_c(0)R_N = \pi\Delta(0)/(2e)$ for an SIS trilayer [8] with $\Delta(0)$, the superconducting gap energy at $T = 0$ is an important figure of merit for device applications. It characterizes the strength of the Cooper pair tunneling since the gap magnitude $\Delta(0)$ is strongly influenced by proximity effects at the barrier–electrode interface. First Josephson junctions with refractory superconductors like Nb and NbN used their natural oxides as an insulator and the sample quality was not high due to the oxygen diffusion into the electrodes and, as a result, appearance of a degraded near-interface region with lower $T_c$. The transition from native oxide barriers to artificial ones preventing the oxygen diffusion [9,10] was a great success towards fabrication of all-refractory Josephson junctions. The bulk $\Delta_{Nb}(0)$ magnitude in ideal Nb samples is near 1.4 meV and hence, theoretically expected $I_c(0)R_N$ product in Nb-AlO$_x$-Nb planar trilayers should be about 2.1 mV. Experimentally, values above $I_c(0)R_N = 1:5$ already indicate a good tunneling barrier [14, p. 21]. At the moment, they typically range between 1.7 and 1.8 mV [13, Table 3.3.1.2]. In NbN-based junctions with MgO artificial barriers, $I_c(0)R_N$ products of 3.2 mV achieved now well agree with the assumed theoretically value of 3.55 mV [13, Table 3.3.1.2].

Such good agreement has been achieved only for conventional low-$T_c$ superconductors. In high-$T_c$ superconductors, the discrepancy between measured and calculated characteristics is much stronger. For example, at 4.2 K $I_cR_N$ products of [001]- and [100]-tilt YBCO bicrystal junctions were found to be only 1–2 mV and several to 10 mV, respectively, those for step-edge and ramp-edge junctions are of several millivolts as well [15]. Most probably, it is due to the peculiar physical properties of cuprates, quasi-two-dimensional materials with electrons moving within copper-oxide ($CuO_2$) layers and *d*-wave symmetry of the superconducting order parameter. Even in $MgB_2$, less anisotropic high-$T_c$ superconductor with an *s*-wave symmetry of the order parameter, reported *I–V* curves are still far from the ideal one, suggesting degradation of the order parameter near the superconductor–barrier interface [15]. As for the recently discovered novel class of high-$T_c$ materials, iron pnictides, the best up-to-date characteristics voltages $I_cR_N$ are of the order of 1 mV [16]. It is not clear if it is a technological problem or its origin is of a fundamental cause which can be related, for example, to the specific $s_\pm$ symmetry of the order parameter.

In the following, we are discussing only low-$T_c$ superconducting devices which are expected to follow a simple relation (2) between $I_c(T)R_N$ and $\Delta(T)$. Notice that even in these samples the statement concerning the $I_c(0)R_N$ product as an indicator of the junction quality is valid only when the barrier transmission probability $D$ is much less than unity. Otherwise, significant contribution to the near-zero voltage bias can arise from the quasiparticle transport. Its presence reveals itself as an excess current $I_{exc}$ given by the intersection of the *I-V* curve slope at high voltage biases with the current axis.

*Temperature stability*. The need of the temperature stability means that small temperature variations should cause only small supercurrent changes [17]. Related temperature dependence of the product $I_c(T)R_N$ for a symmetric SIS trilayer was derived by Ambegaokar and Baratoff [18] who applied the microscopic theory to a tunnel junction geometry:

$$I_c(T)R_N = \frac{\pi\Delta(T)}{2e}\tanh\left(\frac{\Delta(T)}{2k_BT}\right). \tag{2}$$

Temperature dependence of the $I_c(T)R_N$ product in Josephson devices based on conventional superconductors well agrees with this formula [19] whereas it does not apply to high-$T_c$ superconductor junctions, see, for example, the recent overview of experimental data for iron pnictides [16].

From Eq. (2) it follows that conventional refractory-metal SIS junctions exhibit an excellently stable temperature range from 0 to $0.6T_c$ with a huge $I_c$-vs-$T$ dependence above it. Since we are discussing Nb-based junctions at and above 4.2 K, we need the temperature stability at $T \geq 0.5T_c$. As was suggested by us earlier [20], this property can be characterized by an absolute value of the normalized temperature derivative $|d(I_c(T)/I_c(0))/d(T/T_c)|$ - as smaller it is, as less the critical current changes are for a fixed temperature perturbation. We shall return to this issue later by discussing the optimal solution for a self-shunted Josephson junction.

## 3. SELF-SHUNTED JOSEPHSON JUNCTIONS WITH HOMOGENEOUS WEAK-LINK INTERLAYERS

*SNS junctions*. The previous discussion shows that the subgap resistance $R_{sg}$ should be suppressed in order to get rid of the double-valued *I-V* curves shown in Fig. 1a. The simplest way to realize it consists in the replacement of an insulating weak link in the conventional SIS geometry with a normal-metal (N) one whose resistance $R_N$ is much lower than $R_{sg}$ in the tunnel-junction case. Zero-temperature Josephson current through a planar SNS junction in the thick-barrier limit was studied within a stepwise off-diagonal potential approximation by Ishii [21]. It was found that the supercurrent magnitude $I_c$ is inversely proportional to $d_N$ and, in contrast to the sinusoidal supercurrent-phase relation in SIS trilayers [8], its dependence on the difference of two macroscopic wavefunction's phases φ in superconducting electrodes forming an SNS Josephson device is a piecewise linear function with the 2π periodicity. The latter finding is nothing more than the well-known quantum-mechanical statement about the proportionality of the superfluid velocity to the phase gradient of the condensate wave function. This result was confirmed by Svidzinsky [22] who showed *inter alia* that at $T \ll T_c$, the forward-skewed $I_c$-vs-φ relation is transformed into the sinusoidal one with decreasing the transmission probability $D$ of the weak link.

Two seminal papers by Kulik and Omelyanchouk [23,24] proved to be an important impetus for the development of SNS devices. In their first paper [23] the authors considered a short diffusive quasi-one-dimensional wire connecting two S electrodes when the supercurrent can be calculated using Usadel equations. After that, in the framework of Eilenberger equations, they studied a short clean weak link, a constriction with a size in normal and transverse directions less than the electronic mean free path [24] that was found to generate a forward-skewed supercurrent-vs-phase relation at low temperatures. In both cases, the $I_c(0)R_N$ product turned to be larger than that for SIS devices (2) with the factors 1.32 and 2 in dirty [23] and clean [17] limits, respectively. The temperature $I_c(0)R_N$ dependence was also quite different from that for SIS devices predicted by Ambegaokar and Baratoff (2). Similar to the SIS case, it falls quickly with increasing $T$ in the temperature range $T \geq 0.5T_c$, see, for example, Fig. 6.2 in [8].

After the theoretical works cited above, superconductive trilayers with a 10-20 nm thick normal film, where superconducting correlations were mediated by phase-coherent Andreev reflections at the NS interfaces, started to be considered as prospective candidates for Josephson-junction circuitry. Due to the suppressed McCumber-Stewart damping parameter, the SNS junctions revealed inherently non-hysteretic current-voltage characteristics. Moreover, they also exhibited a capacitance intrinsic to the geometry (in the most cases, extremely small) as well as a frequency-dependent reactive component of the material's dielectric response and thus could be incorporated into the RCSJ-like model [25]. The authors of Ref. 26 suggested to name heterostructures with a weak link that contains itself the desired low-resistance component *self-shunted* (or *internally shunted*) junctions.

Unfortunately, the Cooper pair leakage into the normal metal in SNS devices affects the superconductor as well (the latter phenomenon is known as *inverse proximity effect*). Due to the inverse proximity effect depending on the degree of the mismatch between electronic parameters in S and N metals and in contrast to SIS devices, quasiparticle excitations from a normal interlayer penetrate into the junction electrodes, causing suppression of superconductivity near SN interfaces. When normal metals like Au, Ag, Cu are used, then the magnitude of the suppression in the dirty case can exceed two orders of magnitude at $T \approx 0.5T_c$ [12]. In the clean case and at low temperatures $T \ll T_c$ the value of the superconducting order parameter near the NS-interface is equal to nearly half of its value far from the NS-interface [27]. Suppression of the $I_c(0)R_N$ value leads to the reduction in the high-frequency cutoff by orders of magnitude compared to a classical SNS junction [12]. As a result, "most implementations of SNS junctions failed to show an advantage over shunted Nb/Al$_2$O$_3$/Nb tunnel junctions, mainly due to their low $I_cR_N$ product" [26]. One of the ways to get higher $I_c(0)R_N$ values for SNS samples is evidently to use highly resistive weak-link materials.

*Double-barrier SINIS junctions*. Before we assumed an infinitely small interfacial resistance. i.e., electronic characteristics of N and S films were assumed to be not very different from each other. But it is not the case of a highly resistive weak link. For example, when a metal comes in close contact with a semiconductor, a Schottky barrier can arise at their interface [28].

Another possibility discussed below is an artificial barrier between N and S films, for example, an extremely thin oxide interlayer.

When a clean weak link between superconducting electrodes consists of two insulating ultra-thin films and a normal interlayer, the transparency is strongly influenced by interference between normally reflected electrons inside the double-barrier INI region (Breit-Wigner resonances). It means that an appropriate choice of the barrier heights and the normal-metal thickness can provide a weak-link transparency comparable with that of a single N interlayer in SNS samples while superconducting properties in S electrodes will be protected (at least, partly) from the proximity-effect degradation due to the presence of I barriers [29]. As was shown by various theoretical approaches, the physics of a supercurrent via a single Breit-Wigner resonance is similar to that of resonant tunneling via a localized state in SIS tunnel junctions, see Ref. 12 and related references therein. In the latter case, the maximum of the transmission probability is achieved when two barriers are identical and the resonance state is located near the central point of the distance between the two S electrodes [30,31].

The dirty limit of the superconducting transport across the double-barrier weak link was analyzed in detail by Kupriyanov *et al* in their paper [29]. The authors showed that in the case of a short INI weak link the $I_cR_N$ product is determined by two terms. The first one describes direct coupling of the superconducting electrodes, while the second term characterizes the contribution of a sequential tunneling effect in two single-barrier junctions in series. The interplay between the two channels depends on the barrier parameters, temperature, and other factors [29]. It defines also the temperature dependence of the critical current which in some cases can be less sloping, see Fig 20 in Ref. 12.

In principle, SINIS devices could provide solution of the self-shunting problem [32-33]. Unfortunately, due to different starting conditions for the first and the second insulating layers in SINIS samples, the formation of the second barrier is not well controllable in the case of high critical current densities and it leads to increasing barrier asymmetry and the loss of SINIS benefits [34].

*Weak links tuned near the metal-insulator transition*. As was stressed above, highly resistive weak-link materials in SNS junctions could diminish the influence of an N interlayer on S electrodes. In particular, it could be a material near the metal-insulator transition whose resistivity is tuned by adjusting the stoichiometry. Following the paper [26], we can estimate resistivity of such barrier needed to suppress the inverse proximity effect. To avoid the noise impact in 4 K operation, the critical current of the junction should be at least 100 $\mu$A. Since the $I_cR_N$ product is inversely proportional to the width of the SFQ pulse, its value is required to be larger than 320 $\mu$V in order to attain circuit speeds of 50 GHz operation [35]. Hence, the barrier resistance should be, at least, of the order of several Ohms. For $\mu$m-sized SNS junctions we need the resistivity of the order of tens mOhm·cm (compare with $10^{-5}$ mOhm·cm for silver at 4.2 K [36]).

Epitaxial NbN-Ta$_x$N-NbN sandwiches deposited on the lattice-matched MgO substrate were proposed for this purpose in Ref. 37. The weak-link resistivity was tuned to a value near the metal–insulator transition and carefully controlled. For a 13-nm thick barrier the samples exhibited an $I_c R_N$ product greater than 1 mV at 4.2 K. Unfortunately, next investigations [26] showed large variations in $I_c$ and $R_N$ values across a wafer, presumably as a result of variations in the Ta$_x$N stoichiometry and the resulting changes in the barrier resistivity requiring significant improvements in the fabrication process became evident. Another poor conductor near the metal-insulator transition, niobium silicide, has been successfully applied in Josephson circuitry at NIST [38]. It was found that Nb-Nb$_x$Si$_{1-x}$-Nb heterostructures have a wide range of tunability of electrical parameters through control of both the barrier composition and thickness [39-40].

## 4. SELF-SHUNTED JOSEPHSON JUNCTIONS WITH STRONGLY DISORDERED WEAK-LINK INTERLAYERS

**Theoretical background.** Versions of self-shunted Josephson junctions discussed in the previous section were based on the use of homogeneous (or nearly homogeneous) weak-link interlayers. Now we shall consider completely opposite situation when transmission characteristics of a weak link are strongly disordered in the interface plane. The reason for such statement is that a standard Al-oxide tunnel barrier is formed through *in situ* oxygen diffusion into an Al wetting layer and thus includes a lot of internal defects and pinholes as was argued in recent papers [41-43]. The charge flow across such an interlayer is rather a transport through a large number of separate, actually one-dimensional paths than a uniform current across a device cross-section.

Following this hypothesis, we suppose that the main part of the transition region between two superconducting electrodes has very low transmission coefficient $D \ll 1$ while a very small portion of the interface is well transparent with D ≤ 1. Moreover, the 'open' channels are distributed more or less uniformly in the form of filaments having a diameter much less than the superconducting coherence length $\xi_S$ in the junction electrodes whereas the distance between them exceeds $\xi_S$. In this case, the inverse proximity effect on the S layers should be very small and the superconducting order parameter (even near the NS interface) remains almost the same as in the bulk. The supercurrent that flows through the low-transparent (and thus tunnel-like) part of the weak link will follow the Ambegaokar-Baratoff theory [18] whereas the transport of Cooper pairs across high-transparent filaments will realize internal shunting following the SNS scenario developed by Kulik and Omelyanchouk [23,24].

Mathematically, the distribution of a random variable with two distinct peaks (local maxima) is known as a bimodal distribution. In practical Josephson devices, there is a huge number of transmission channels $N \gg 1$ with different individual transparencies $D_i$ where $i = 1, \ldots, N$. Inhomogeneous distribution of the channel transmission probabilities can be proven by experiments on few-mode junctions that were fabricated in very similar conditions and thus should be identical from the first sight. Here we refer to a recent work [44] where supercurrent-phase relations in mesoscopic Josephson junctions made of gate-tunable semiconductor InAs nanowires with superconducting Al leads were measured at ultra-low temperatures. Unexpectedly, a vast majority of forward-skewed supercurrent-vs-phase relations agreed with the Kulik-Omelyanchouk prediction [24] for SNS junctions. Such behavior consistent with a resonant-tunneling effect [44] could be attributed to the charge transport via an isolated localized state with an energy $\varepsilon_{res}$ and level widths $\Gamma_{L(R)}$ which arise due to the decay into the left (right) lead and exponentially depend on the distance $\delta$ of the impurity from the middle of the barrier. The elastic-transmission probability $D$ in this case is given by the Lorentzian-like formula [45,46]

$$D = \frac{4\Gamma_L \Gamma_R}{(\varepsilon - \varepsilon_{res})^2 + (\Gamma_L + \Gamma_R)^2} \,, \tag{3}$$

where $\varepsilon$ is the energy of an incident electron. Frequent realization of a middle-located state with $\Gamma_L = \Gamma_R$ and hence $D \leq 1$ is not clear but more likely is of an intrinsic origin nature. From Eq. (3) it follows that the resonant transmission regime can be destroyed not only by the asymmetry in the tunnel rates $\Gamma_{L(R)}$ but also by the shift $\delta$ of a resonance state from the middle of the weak link. In the latter case, we get $D = \cosh^{-2}(\delta / a_0)$ for $\varepsilon \approx \varepsilon_{res}$ where $a_0$ is the localization radius.

Now we shall discuss another physical realization, a planar symmetric SIS trilayer with the wave-function decay length $\kappa^{-1}$ in the barrier and show that on some cases we can also expect a significant (although not a dominant) part of the high-transparent eigenchannels in a strongly disordered multi-mode weak link. The transparency $D$ of such junction is given by the well-known relation [47[

$$D = \frac{4k_x^2 \kappa^2}{(k_x^2 - \kappa^2)^2 \sinh^2(\kappa d) + 4k_x^2 \kappa^2 \cosh^2(\kappa d)}, \tag{4}$$

where $k_x$ is the wave-vector component in the metallic electrodes normal to metal-insulator interfaces, $d$ is the barrier thickness. Let us simplify the relation (4) starting with the case of an ultrathin potential barrier with a constant height that is extending from $x = 0$ to $x = d \ll \kappa^{-1}$ and consider two limits, a high barrier with $\kappa \gg k_x$ and a low one with $\kappa \ll k_x$. In both cases, we get a Lorentzian $D = 1/(1+Z^2)$ with the governing parameter $Z$ equal to $\kappa^2 d/(2k_x)$ and $k_x d/2$, respectively [48]. We can expect that the Lorentzian approximation for the transmission coefficient Z is appropriate in the intermediate instances as well. Now we assume a slowly varying potential barrier when a semiclassical WKB approximation is applicable to the transition region between two metallic electrodes. Then, at classically turning points $x_L$ and $x_R$ $k_x = \kappa(x_L) = \kappa(x_R)$ and Eq. (4) reduces to the relation $D = \cosh^{-2}\left(\int_{x_L}^{x_R} \kappa(x) dx\right)$.

Let us transfer from transparencies of individual transmission channels to a Josephson junction with a comparatively large area where the total transmission coefficient across a weak link is defined by a sum of independent local transparencies $D_i$ for concerned eigenchannels. It has been shown above that under certain conditions we can expect one of two simple dependences of the single channel conductance $G$ on dimensionless quantities $Z$ and $Y$, $G(Z) = G_0(1+Z^2)^{-1}$ and $G(Y) = G_0 \cosh^{-2}(Y)$ with the conductance quantum $G_0 = 2e^2/h$. Assume that $Z$ and $Y$ are random variables uniformly distributed from zero to infinity. Then $\rho(Z) = 2\hbar \bar{G}/e^2 = \text{const}$ in the first case and $\rho(Y) = \pi\hbar \bar{G}/e^2 = \text{const}$ in the second case (the disorder-averaged macroscopic conductance $\bar{G}$ was introduced to normalize $\rho(Z)$ and $\rho(Y)$ so that $\bar{G} = \int_0^\infty \rho(Z)G(Z)dZ$ and $\bar{G} = \int_0^\infty \rho(Y)G(Y)dY$). With the parametrization $D = (1+Z^2)^{-1}$ we can transfer from $\rho(Z)$ to the distribution of local transparencies $D$ $\rho(D) = \rho(Z)(dZ/dD)$. For resonant tunneling via middle-localized isolated states and a rectangular-like ultrathin barrier, two cases discussed above, it reads

$$\rho_{SB}(D) = \frac{\hbar \bar{G}}{e^2} \frac{1}{D^{3/2}(1-D)^{1/2}} . \qquad (5)$$

Similar analytical formula can be obtained following $\rho(Y)$ dependence valid for resonance states spatially shifted within the barrier and slowly varying potential barriers

$$\rho_{D}(D) = \frac{\pi \hbar \bar{G}}{2e^2} \frac{1}{D(1-D)^{1/2}} . \qquad (6)$$

Analytical relations (5) and (6) were earlier obtained by Schep and Bauer [49] for extremely high and infinitely thin barriers between metallic electrodes and by Dorokhov [50] for diffusive conductors, respectively. It follows from above that applicability of these relations is much broader than assumptions under which they were first derived and, hence, they may be universal. In this context, the notion of universality means that quantitative features of the charge transport across a heterostructure with a locally inhomogeneous barrier are not sample specific and can be deduced from a single global parameter, without requiring knowledge of the system details.

We should emphasize that both distribution functions are just bimodal with a significant amount of "open" channels for which $D \leq 1$ and that the difference between them is comparatively small [46]. Note also that in both cases transport characteristics of the metal-weak link-metal trilayers are controlled by the only macroscopic quantity $\bar{G} = \int_0^1 \rho(D) G(D) dD$, the disorder-averaged macroscopic conductance, where $G(D) = (2e^2/h)D$ [51]. From the previous discussion, it follows that preconditions for the realization of the universal distribution functions are (i) a relatively complex structure of the weak link between two metallic electrodes, (ii) the same functional dependence of a transmission coefficient on a single random variable for all eigenchannels, and (iii) an evenly spread of the variable over all admissible values.

It is noticeable that before the work [49], the same bimodal distribution (5) was derived for a quasi-ballistic double-barrier INI interspace with two identical uniform insulating I layers [52], a system that physically is very different from a thin disordered dielectric film. As was explained in [53], coincidence between the two systems occurs due to the fact that, for a finite thickness of the N interlayer and symmetrical NIN structure the transmission coefficient $D$ is also of a Lorentzian-like form with an injection angle between the wave vector of an incident charge and the interface normal as a controlling parameter.

**Experimental verification.** The bimodal distribution function is able to provide a solution of the self-shunted problem. The main question is how to implement the three preconditions formulated above without fundamental changes to the existing technology. Before to answer this question, we should propose the way for the experimental verification of the reliability of universal distribution functions. We shall show that the adequate normalization of experimental curves measured in the superconducting state permits to eliminate the only adjustable parameter $\bar{G}$ and to test the feasibility of a bimodal transparency distribution in Josephson junctions *without any fitting parameters.*

The first class of intrinsically shunted Josephson junctions proposed by us [54] are Nb/Al-AlO$_x$-Nb Josephson junctions fabricated using a standard Nb-technology but with an insulating barrier thinner than that in conventional tunnel Nb-AlO$_x$-Nb trilayers [9,10], see the inset in Fig. 2. Such a junction may be regarded either as a modification of a high-current-density Nb-AlO$_x$-Nb junction [55] with an additional Al interlayer between one of Nb electrodes and the Al oxide, or as a strongly asymmetrical S-I$_1$-N-I$_2$-S junction [34] where I$_2$ is the Al-oxide tunnel barrier and I$_1$ represents the Nb/Al interface of a finite transparency. The aim of introducing the Al interlayer with the finite thickness $d_{Al}$ has been two-fold, first, to protect the junction against the impact of pinholes in the ultra-thin AlO$_x$ barrier and, second, to improve the temperature stability of the supercurrent at the operating temperatures at and above 4.2 K, see below.

As was emphasized in Refs. 53 and 56, superconducting properties of our S/N-I-S structures with $d_{Al}$ ranging from 40 to 150 nm are governed by the only controlling parameter $\gamma_{eff} \approx \gamma_{Nb/Al} d_{Al} \sqrt{T_c^{Nb}/T_c^{Al}} / \xi_{Al}$ where $\gamma_{Nb/Al}$ is the reduced S/N interface resistance that follows a phenomenological relation for Nb/Al bilayers $\gamma_{Nb/Al}(d_{Al}) = 0.111 d_{Al}$, $d_{Al}$ being in nm, $T_c^{Nb}$ and $T_c^{Al}$ are related critical temperatures, $\xi_{Al}$ is the superconducting coherence length in Al that changes from 110 nm in our thinnest Al films to 150 nm in the thickest ones [57]. Thus, for fixed fabrication conditions the ratio of $d_{Al}$ to $\xi_{Al}* = \xi_{Al}\sqrt{T_{c,Al}/T_{c,Nb}} = 50 \div 60$ nm determines the spatial dependence of the superconducting order parameter in our Nb/Al-AlO$_x$-Nb samples.

When $d_{Al}$ is much less than $\xi_{Al}*$, the system studied can be interpreted as an asymmetric S$_1$IS junction where S$_1$ is the S/N bilayer. Proximized S=Nb layer induces superconducting correlations in a very thin N=Al film which can be considered spatially independent. For a known distribution function $\rho_{SB}(D)$ applicable to disordered ultra-thin barriers, the quasiparticle current-vs-voltage curve of the S$_1$IS device can be written as a sum of independent contributions from eigenchannels $I_{qp}(V) = \int_0^1 dD \rho(D) I_{qp}(D,V)$. Since a part of the modes belong to "open" channels, their individual $I_{qp}(D \leq 1, V)$ characteristics should be calculated taking into account multiple Andreev electron-into-hole (and inverse) reflections in the near-barrier interspace [58,59]. The energy of an electron-like quasiparticle transferring the barrier is increased by $eV$ each time when it crosses it, for example, from left to right (and from right to left for a hole-like excitation). These scattering events will continue back and forth and with each round-trip the energy of an electron-like quasiparticle will increase by the $2eV$ value. If so, the transport across the S/N-I-S device can be described in terms of Andreev-reflection amplitudes $a_1(\omega) = i\left(\omega - \sqrt{\omega^2 + \Phi_{Al}^2(\omega)}\right)/\Phi_{Al}(\omega)$ from the Al interlayer where $\Phi(\omega) = \omega F(\omega)/G(\omega)$ is the ratio of modified and normal Green's functions $F$ and $G$ and $a(\omega) = i\left(\omega - \sqrt{\omega^2 + \Delta_{Nb}^2}\right)/\Delta_{Nb}$, $\Delta_{Nb}$ is the energy gap of Nb. The simplest approximation for the function $\Phi_{Al}(\omega)$ in the Nb/Al bilayer looks as $\Phi_{Al}(\omega) = \Delta_{Nb}/(1 + \gamma\sqrt{\omega^2 + \Delta_{Nb}^2}/\Delta_{Nb})$ [60] with a fitting parameter $\gamma$ that can be found by using the gap magnitude in the Al interlayer found experimentally [54]. When $d_{Al}$ is much more than $\xi_{Al}^*$, the Al film can be considered non-superconducting since the two layers become more detached by the increased $\gamma_{eff}$ value. Then the main proximity-induced changes are originated from the Andreev reflections at the S/N interface accompanied by phase shifts arisen from electron (a hole) passages across the N layer. A final mean free path in the Al film caused by impurity scatterings can be introduced into this scheme, as explained in Ref. 56.

Now, using the distribution function (5) for ideally disordered insulating barrier, we can calculate dissipative current-voltage characteristics for the four-layered S/N-I-S devices in both cases. At very low voltages $V << \Delta_{Nb}/e$ the *I-V* dependence is linear $I_{qp}^{(S)}(V) = V/R_{sg}$ as well as in the normal state $I^{(N)}(V) = V/R_N$ (Fig. 2). The ratio of the two resistances does not depend on any parameter and thus can serve for the experimental verification of the assumptions relating the self-shunted nature of the junctions. Subgap ohmic resistance $R_{sg}$ was extracted from experimental data as the slope of a best-fit linear regression line for $I_{qp}^{(S)}(V)$ curves in the interval from 0 to 0.2 mV (the main panel in Fig. 2). The normal-state resistance $R_N$ was determined from a linear fit to dissipative current–voltage curves at ~ 1mV (see Fig. 2) or after suppressing superconductivity by external magnetic fields. Numerical and experimental results for the $R_{sg}/R_N$ ratio at 1.7 K are compared in the inset in Fig. 2. The agreement is good (most probably, superconducting correlations induced in the Al interlayer due to the proximity to a Nb electrode are weaker than those predicted in the dirty limit).

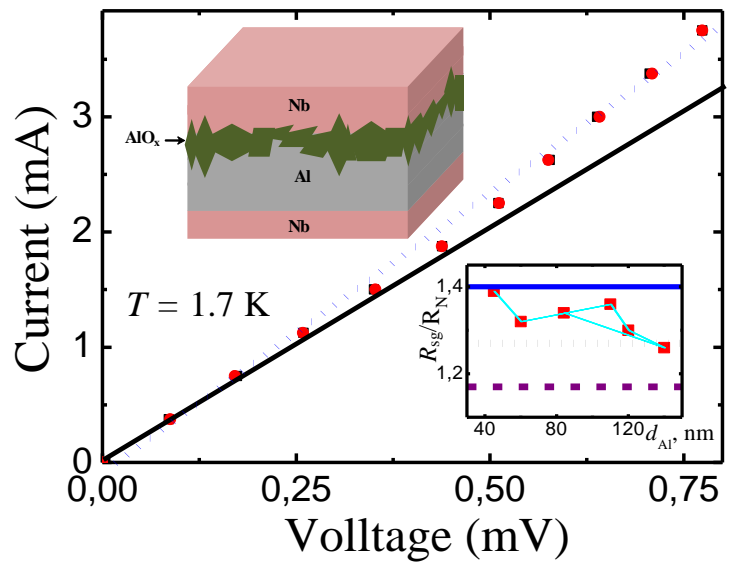

*Fig. 2.* Quasiparticle (dissipative) current–voltage characteristic (squares) of a representative Nb/Al–AlO$_x$–Nb junction shown schematically in the left inset, $d_{Al}$ = 140 nm, $T$ = 1.7 K. Solid and dotted straight lines correspond to Ohm's laws with $R_{sg}$ and $R_N$ resistances, respectively. The right inset shows $R_{sg}/R_N$ ratios for Al films with different thicknesses $d_{Al}$; solid, dotted, and dashed lines correspond to theoretical $R_{sg}/R_N$ values for N-I-S, S/N-I-S (quasi-clean limit [56]), and S/N-I-S (dirty limit with proximity-effect account [53]), respectively.

The second class of intrinsically shunted Josephson junctions are those with low-height and, hence, comparatively thick interlayers of strongly disordered semiconductors with metallic droplets inside them. Transport across such heterostructures can be dominated whether by tunneling via those configurations of localized states in the semiconductor layer that permit resonant transmission of electrons, see Eq. (3) [61] schematically shown in the left inset in Fig. 3 as extremely thin filaments connecting two sides of the Josephson junctions. In this case, some kind of a bimodal distribution may be realized as well. A large number of "open" eigenchannels with the transparency $D \leq 1$ would reveal themselves, in particular, in the emergence of an excess current $I_{exc}$, a constant shift of the superconducting *I-V* curve towards that measured in the normal state at *V* exceeding $\Delta_S/e$, see the main panel in Fig. 3. In the right inset in Fig. 3 we compare calculated and measured values of the ratio $I_c/I_{exc}$ at 4.2 K [62] and show that it provides a second way to verify the validity of a universal distribution function. Note that for a superconducting junction without a barrier ($D$ = 1) $I_c/I_{exc} \approx$ 1.2 and 1.3 at 0 and 4.2 K whereas in the tunneling limit ($D \ll 1$) $I_{exc} \to 0$ and, hence, $I_c / I_{exc} \to \infty$. Averaging the formula for $I_{exc}$ with the distribution function (5) we get $I_c / I_{exc} \approx 1.7$ and 2.4 at 0 and 4.2 K [56].

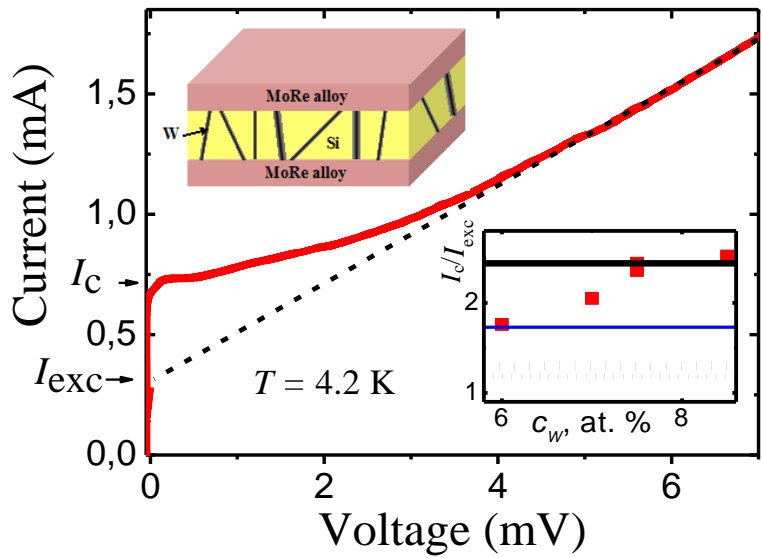

*Fig. 3*. Current–voltage characteristic (solid line) of a representative MoRe–Si:W–MoRe junction shown schematically in the left inset, the dopant concentration $c_W = 7.5$ at.%, $d_{Si} = 38$ nm, $T = 4.2$ K, a high-voltage asymptote (dashed line) exhibits the presence of an excess current $I_{exc}$ [62]. The right inset shows $I_c/I_{exc}$ ratios for different dopant concentrations $c_W$, thick solid and dotted lines correspond to theoretical $I_c/I_{exc}$ values for a strongly inhomogeneous barrier with a universal distribution function (5) and without it, respectively, calculated at 4.2 K, whereas related thin lines are for 0 K.

In the right inset in Fig. 3 we present theoretical expectations [56] and related experimental data for five superconductive junctions formed by superconducting MoRe-alloy electrodes and a several tens nm thick Si interlayer doped by tungsten [62]. From the comparison in Fig. 3 it follows that the ratio $I_c/I_{exc}$ increases with the dopant concentrations $c_W$ due to the enhancement of the number of transport channels.

**Sinusoidal current-phase relation and temperature stability at 4.2 K.** Operation of superconducting devices at and above 4.2 K allows the usage of cheaper and more compact refrigeration systems. The main three aspects of the proposed SNIS devices that are important at temperatures above $0.5T_c$ are as follows: (i) hysteretic at very low temperatures DC current–voltage characteristics are converted into single-valued ones [54], (ii) complicated current-phase relation at $T \ll T_c$ is transformed into the standard sinusoidal form, leaving only harmonic current–voltage steps under microwave irradiation [17], and (iii) the temperature stability strongly increases just above $0.5T_c$ [20].

In the Nb/Al-AlO$_x$-Nb junctions, at $T \ll 0.5T_c$, we have a set of transport eigenchannels with different $I_c$-vs-$\varphi$ relations from forward-skewed ones inherent for open eigenchannels [24] till standard sinusoidal dependences for nearly closed channels. The presence of the additional normal Al interlayer complicates the situation since in this case the $I_c$-vs-$\varphi$ dependence contains more than one sinusoidal term with different periods (see the related discussion in [20]). Fortunately, the current-phase relation in the SNIS structures with a relatively thick normal layer exhibits a pure harmonic behavior when the operating temperature exceeds $0.5T_c$ [17].

Temperature dependence of the critical supercurrent in the discussed SNIS devices essentially depends on the interrelation between the N-interlayer thickness $d_N$ and its coherence length $\xi_N$. In Ref. 53 we discussed two related models. The first one, valid for $d_N \ll \xi_N$, [54] is based on the dirty-limit approach [29] that describes stationary properties of double-barrier S-I$_1$-S′-I$_2$-S junctions with arbitrary resistances of I$_1$ and I$_2$ interlayers (in our case Nb/Al interface and the insulating AlO$_x$ layer, respectively). The simulated $I_c$–vs–$T$ curves shown in Fig. 4 are governed by a single fitting parameter $\gamma_{eff}$ introduced above. The second theoretical model valid for $d_N \gg \xi_N$ [63] assumes specular scatterings at the S/N and N-I interfaces and is controlled by a parameter $\alpha = 2d_N \Delta_S / (\hbar v_N)$, where $v_N$ is the Fermi velocity in the N interlayer. The $\alpha$ value determines positions of Andreev bound states formed within the energy gap in the N interlayer.

It follows from Ref. 64 that the shape of a measured $I_c$-vs-$T$ dependence for an SNIS device can provide significant information concerning the N-film conducting state as well as that of the Nb/Al interface - whether it is dirty and the scatterings at the interface are diffusive or comparatively clean and the scatterings are specular. In the first case, the $I_c$-vs-$T$ curves are concave upward due to the induced energy gap in the dirty normal side of the S/N bilayer while in the second case, they are concave downward and the charge transport occurs via Andreev bound states in the clean N interlayer. Our experiments on overdamped Nb/Al-Al oxide-Nb devices have revealed mainly the first type of the superconducting characteristics and thus indicate the presence of a minigap in the samples, see Fig. 4. Due to the latter circumstance, $I_c$-vs-$T$ dependences become flatter above $0.5T_c$ with increasing $d_N$. It means that the temperature stability will increase, see Fig. 4, and a compromise between suppression of the critical supercurrent and enhanced temperature stability will allow to realize a superconductive heterostructure with optimal characteristics required in each particular case.

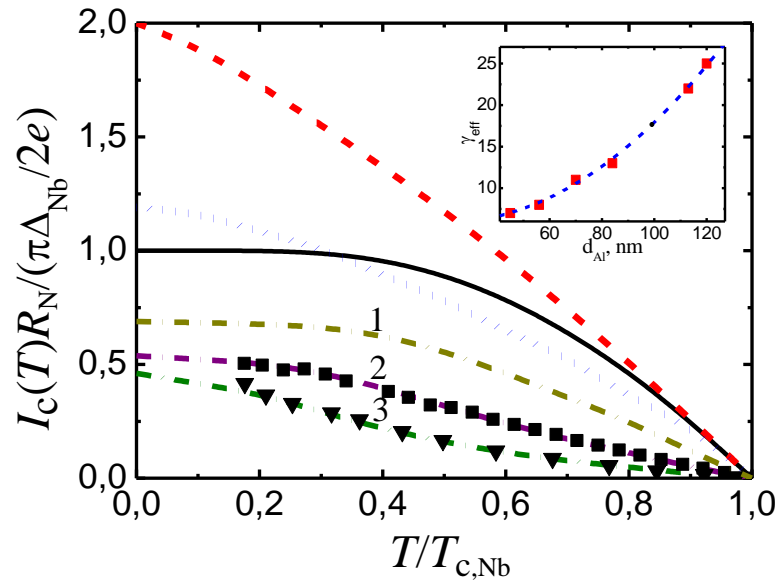

*Fig. 4*. $I_c$-vs-$T$ curve simulations for standard SIS [18] and SNS [24] Nb-based trilayers (solid and dashed curves, respectively) compared with the related data for Nb/Al–AlO$_x$–Nb junctions where the distribution function of local barrier transparencies is described by Eq. (5), the dotted curve is for $d_{Al} = 0$ and the dashed-dotted curves are for finite $d_{Al}$. The curves 1, 2, and 3 were calculated for $\gamma_{eff}$ = 2, 8 ($d_{Al} \approx 56$ nm), and 22 ($d_{Al} \approx 113$ nm), respectively. One-to-one correspondence between the controlling parameter $\gamma_{eff}$ and $d_{Al}$ follows from the empirical relationship given in the inset. Experimental data in the main panel and the inset are shown by symbols.

## 5. CONCLUSIONS

The energy dissipation of CMOS transistors is reaching physical limits and has become an important barrier to building more powerful supercomputers [1,2]. Digital integrated circuits based on superconductive SFQ logic offers a combination of high-speed and ultralow power dissipation unmatched by any other device [3]. Hence, superconducting computing can be considered as a potential solution for further progress in this field. The most common devices remain to be Nb-AlO$_x$-Nb underdamped SIS junctions [9,10] with good uniform and reproducible properties [64]. On this way, one of the main problems is the need for self-shunted Josephson junctions [5]. There is significant progress with non-hysteretic SNS Josephson junctions of the NbN-Ta$_X$N-NbN type [65] and the Nb-Nb$_x$Si$_{1-x}$-Nb type [40] where the barriers are tuned at the metal-insulator transition. Internally shunted SNIS junctions based on the Nb/Al-AlO$_x$-Nb technology can be used for a large-scale integration in superconductive circuitry as well [66].

Another research area which requires self-shunted devices is a quantum voltage standard, a complex system that applies a superconductive integrated circuit chip to generate stable voltages that depend only on fundamental constants and an applied frequency. These features known as Shapiro steps appear due to the synchronization between the external driving frequency $f$ and the Josephson-junction intrinsic frequency [8]. The accuracy and universality of the relation between their voltage positions and $f$ has made the Shapiro steps the basis of the international voltage standard with an accuracy at least of one part per billion [67]. However, the extension of this accuracy to AC voltage standards is not able with hysteretic devices since strongly overlapping voltage steps make it difficult to adjust rapidly a certain voltage for the synthesis of a waveform with the needed accuracy. To overcome this limit, series arrays of overdamped Josephson junctions with non-hysteretic DC characteristics should be used. The main prevalent tendency at the moment is to replace the metallic interlayer with a barrier near the metal-to-insulator/semiconductor transition. It permits to tune the barrier resistivity and the characteristic voltage of the junctions [68]. The most successful example is an SNS-like Josephson trilayer based on a $Nb_xSi_{1-x}$ barrier first proposed at NIST [69] and then developed at PTB [70,71]. SNIS junctions have also proved to be suitable for the realization of a programmable voltage standard. Even moreover, they have showed the possibility of operation at 4.2 K being biased on a higher order step and therefore can potentially reduce the number of junctions on the chip as well as an operation at higher temperatures, up to 6-7 K employing energy efficient cryocoolers [72].

Let us stress again the advantages of Nb/Al-AlO$_x$-Nb multilayers for their usage at 4.2 K and above: (i) well-established Nb/Al technology [9,10] slightly modified for SNIS junctions, (ii) simple self-averaging mechanism due to an ultra-thin strongly disordered Al-oxide interlayer and decreased at 4.2 K supercurrent [54], (iii) prevention of a direct current through a pinhole in a barrier using an additional Al interlayer in the weak link, (iv) averaging of spatial variations of Al superconducting parameters due to the large coherence length in the Al interlayer, (v) reasonably high values of the critical current density and characteristic voltages at 4.2 K [54], (vi) nearly sinusoidal $I$c-vs-φ relation and temperature stability at operating temperatures exceeding $0.5T_c$ [17,63]. It makes the SNIS junctions appealing for superconductive electronics. Advanced studies, including reliability and reproducibility of such samples, are likely to lead to further progress in this field.

Very new way for the controllable creation of internal shunting in a nanometer-thick interlayer in Josephson junctions may be implemented with binary [73] and complex [74-77] transition-metal-oxide weak links through the formation of self-aligned conductive filaments using an electrical resistive-switching process. In our opinion, such manipulation of inhomogenities in weak links between superconducting electrodes has potential to create novel types of self-shunted Josephson junctions and could be a verdant area for future work.

At last, we would like to emphasize once more that the ideas proposed for realizing intrinsically shunted Josephson devices were generated by two papers [23,24] published by I.O. Kulik and A.N. Omelyanchouk in 70s. It is our great pleasure to salute Professor Alexander Omelyanchouk for his upcoming 70th birthday, to congratulate with all seminal works done up to now, and to express our hopes for new significant scientific results in future.